\documentstyle[aps,prl]{revtex}
\begin{document}
%\draft

%\twocolumn[\hsize\textwidth\columnwidth\hsize\csname@twocolumnfalse\endcsname

\title{Selective d-state Conduction Blocking in Nickel Nanocontacts}
\author{A. Smogunov, A. Dal Corso, and E. Tosatti}
\address{SISSA, Via Beirut 2/4, 34014 Trieste (Italy)} 
\address{INFM, Unit\`a di Trieste Via Beirut 2/4, 34014 Trieste (Italy)} 
\address{ICTP, Strada Costiera 11, 34014 Trieste (Italy)} 
\date{\today} 
\maketitle 

\begin{abstract} 
The lowest conductance step for a Ni nanocontact is anomalously small 
in comparison with the large expected number of conducting
channels. We present electronic structure calculations for an extremely
idealized Ni nanobridge consisting of just a monatomic nanowire.
Our calculations show that no less than eight single spin bands
cross the Fermi level in a nonmagnetic Ni monatomic
wire, dropping marginally to seven in the more stable, fully 
ferromagnetic state. However, when we build in the wire a 
magnetization reversal, or domain wall, by forcing the net magnetization 
to be zero, we suddenly find that $d$ electrons selectively cease to
propagate across the wall. $s$ electron propagation remains, and
can account for the small observed conductance steps.

\vskip 1.0truecm
\end{abstract}

\section{Introduction}

Recent conductance data of nanocontacts and break junctions in a 
magnetic transition metals such as Ni have shown interesting and 
partly unexpected results. 
While early conductance histograms for Ni at room temperature in air appeared 
basically structureless \cite{costa}, Oshima et al. \cite{oshima},
who worked in vacuum, at variable temperature, and with the possibility
of a magnetic field, found conductance steps preferentially near 2
and 4 (in units of $g_0=e^2/h$, the conductance quantum per spin)
at RT and zero field, near 4 at 770 K and zero field, and near
3 (occasionally near 1) at RT with a field. Ono et al. \cite{ono}, 
reported again 2 for Ni in zero field, and 1 for Ni in a field. 
Break junction data by Yanson \cite{yanson} show a step 
at conductance about 3.2 in zero field. While the reasons for this 
diversity are  not always clear, there seems to be a consensus, based also 
on other data for noble metals, that the ultimate contact must be
monatomic. For a monatomic noble metal contact, for example, the
availability of a single $s$ electron channel for both spins species 
immediately rationalizes a conductance step of 2, as is generally observed
\cite{costa,ono,yanson}. 

The general question which we broach -- if not yet fully solve of course  --
is what to expect for the number of truly conducting electronic channels 
in a monatomic nanocontact of a magnetic {\it transition} metal, one 
possessing partly filled, bulk-polarized $d$ electron states besides the 
$s$. In  order to pursue this first 
rationalization attempt we shall purposely adopt an oversimplified model, 
consisting of just a single monatomic, regular, tipless Ni 
chain. While that model is surely quite different from the true 
nanocontact, as it neglects the presence of the supporting tips 
and the general lack of regularity near the ultimate bridging atom,
its simplicity is crucially useful. It allows a detailed microscopic 
study, that can bring to light new potential phenomena. 

Because the model possesses no tips, the only measure of conductance we will 
discuss will be simply the number of one-dimensional bands for each
given spin ("channels") 
crossing the Fermi level, no allowance made at this stage for imperfect
transmission, and for the ensuing noninteger values found in reality.
Moreover, standard electronic structure calculations are mean-field
in character, imply for a transition metal one-dimensional nanowire 
a  magnetic long-range order which is in reality suppressed by 
fluctuations, or that can only be supported by tips under special 
spatial circumstances. Nonetheless, since our work is aimed at understanding 
much more basic questions, these problems are not really relevant at 
this stage. The mean field channel counting is moreover probably not
bad anyway, as shown for example by the large U 1D Hubbard model, 
where $K_{\rho} = 1/2$ in that spin-singlet Luttinger liquid 
\cite{schulz} implies one effective conducting channel in place of two, 
exactly the same as if there were magnetic long-range order.
 
Our results indicate that the minimal channel number of the
nonmagnetic monatomic Ni nanowire should be as large as 8, a large
number when compared with experimental steps around 2 and 4 (large even
after considering a limited transmission for a $d$ state). 
Ferromagnetic polarization, which in mean field is quite strong and lowers 
the energy considerably, only reduces the channel number to 7, and
does not remove the disagreement. However, insertion of a single 
magnetization reversal (a sort of collinear "Bloch wall") inside the
monatomic magnetic nanowire leads to a sharp drop from 7 to 2 channels,
a value now much closer to the experiments. Inspection of calculated 
bands and wavefunctions clearly indicates that the magnetization reversal 
along the wire leads to a selective blocking of $d$-electron propagation, 
leaving only the $s$-electron fully conducting.

\section{Method}

Calculations were carried out in the framework of
density-functional theory (DFT) in the local density approximation (LDA). 
The Perdew-Zunger
parametrization \cite{perdew} of exchange-correlation energy was used.
The nuclei and the core electrons are described by ultrasoft 
pseudopotentials \cite{vanderbilt} with the parameters of Ref.~[8].
The Ni single atom contact was simulated by a regular monatomic
Ni wire, infinite along the $z$ axis, and periodically repeated
in a square lattice along $x$ and $y$. The spacing used, 8.46 \AA, was
checked to be large enough to avoid measurable wire-wire interactions.
The kinetic energy cut-offs of the plane wave basis set were 25 Ry
and 300 Ry for the wave functions and the charge density, respectively.
Integration of the 1D Brillouin zone was done on a uniform mesh 
of 80 $k$-points.  These parameters proved sufficient to provide converged
results. The integration up to the Fermi level was done by
a standard broadening technique \cite{smearing} with the smearing 
parameter of 0.002 Ry.

A single Ni atom per cell is required for both nonmagnetic and
ferromagnetic states. To study the state with spin reversal (actually
two spin reversals with periodic boundary conditions) we extended our
cell up to 8 Ni atoms. While our calculations were restricted to collinear
spins thus excluding for simplicity the possibility of spin moment
rotation, the magnetization magnitude and sign were allowed to 
vary with total freedom as a function  of position. In order to 
find the state of lowest energy with a given total magnetization we 
applied the fixed-spin-moment (FSM)
method \cite{fms1,fms2}, introducing two Fermi energies $E_F^\pm$ for different
spin directions. The difference $B=(E_F^+-E_F^-)/2$ is a magnetic field which
is needed to stabilize the state at the chosen magnetization. The stable
configurations  correspond to zero magnetic field (when $E_F^+=E_F^-$).  

\section{Ferromagnetic and nonmagnetic monatomic wires}

The ground state of a monatomic Ni wire in a large interval of interatomic
distances (including the limiting case of infinitely separated free Ni atoms)
is calculated to be ferromagnetic (Fig.~1). The nonmagnetic state
(magnetization everywhere identical to zero) has a higher total energy
compared to the ferromagnetic one, by about 0.12 eV/atom 
at the equilibrium spacing (total energy minimum). The
equilibrium interatomic distances both for nonmagnetic
($a=2.06$ \AA) and for ferromagnetic ($a=2.11 $ \AA) Ni wires are found to
be considerably smaller than the corresponding bulk value $a_{\rm Bulk}=2.42$
\AA. Not surprisingly, magnetism suddenly disappears, and the wire turns
nonmagnetic below a critical spacing of about 1.9 \AA. This situation is 
demonstrated in the inset of Fig.~1 where the magnetic moments of bulk 
Ni and of an isolated Ni atom are also indicated. As could be expected the 
equilibrium moment of the wire  ($M=1.11 \mu_B$) is intermediate between 
the atomic and the bulk values.

The ballistic conductance of a contact is given by Landauer's formula
which in the case of perfect transmission and independent channels
has a simple form $G=Ng_0$, where $g_0=e^2/h$ is the conductance half 
quantum and $N$ is the number of conducting channels per spin. We thus
estimated the total conductance by simply counting the number of channels, 
that is the number of bands of either spin crossing the Fermi level. 
In Fig.~2 we present the band structure of the nonmagnetic and ferromagnetic
Ni wires at their equilibrium interatomic distances. The valence
electrons  in the atom are $3d$ and $4s$, and each band is correspondingly 
labeled by its main atomic character. The  $s$ and $d_{z^2}$ bands are 
strongly hybridized, while $d_{x^2-y^2}$, $d_{xy}$ are split from
$d_{xz},d_{yz}$ by the uniaxial wire crystal field.
There are 8 channels in the nonmagnetic wire,
reduced to 7  in the ferromagnetic wire. That is so much larger
than the basic experimental conductance step of 2, to suggest the need
to identify some mechanism blocking some of the channels -- presumably the 
$d$ channels and allowing only the remaining to conduct.

\section{Nanowire with a domain wall: $d$ electron blocking}

A ferromagnet generally contains Bloch walls, separating domains with different
magnetization directions. Inside the wall, which is generally rather thick,
the magnetization rotates gradually between one direction and the other.
Given two tips connected by a nanocontact, it is possible that a wall might
be trapped precisely there \cite{ono}. Although in that case the spin 
reversal might be more abrupt, and thus more costly per unit section 
than in the bulk, the monatomic contact size could nonetheless minimize 
the total energy cost. We thus investigated the possibility that the
two tips in the nanocontact under some circumstances might be magnetized
in opposite directions forming thereby a spin reversal configuration.
The spin reversal costs some energy so that this configuration should 
have an intermediate energy between the uniform ferromagnetic and the 
nonmagnetic states. To simulate this situation we considered large cell 
monatomic wire calculations 
where any local magnetization was allowed, but the total magnetization $M$
was required to vanish, $M=0$. In a sufficiently long Ni wire, that 
must lead to a ground state consisting of periodically repeated up and 
down spin ferromagnetic segments, separated by walls. Moreover if the
walls are sharp, which turns out to be the case, the cell length can 
be quite small, and we found a unit cell including 8 Ni atoms quite 
adequate. The resulting ground state has, roughly speaking, 4 atoms 
with positive and 4 atoms with negative magnetization 
(Fig.~3a). This state is more stable than the nonmagnetic state, because
the cost of the two walls, estimated to be $\approx 2 \times 0.2$ eV, 
is smaller than the energy gained by magnetizing the two domains, 
roughly $8 \times 0.12$ eV. Moreover, symmetry at $M=0$ requires that 
$B=E_F^+-E_F^-=0$, so that the auxiliary magnetic field of this state 
is zero, and the energy bands (Fig.~3b) are spin degenerate.

The new striking feature is that some of the $d$ bands, including the
$d_{xy},d_{x^2-y^2}$ bands, formerly crossing the Fermi level, have now 
turned into flat dispersionless levels. The wave functions of these bands
(Fig.~4) are strongly localized and therefore cannot 
contribute to conductance.  Viceversa the $s-d_{z^2}$ states which remain
delocalized are still able to conduct. As a result the number of conducting
channels has been reduced by the wall from 7 to 2. 

This selective $d$ electron blocking is easily rationalized in terms of 
the up and down spin effective potentials, which inside each domain 
have a sizable offset, and which alternate as one 
moves from one domain to the next. The $d$ states, whose mass is large, 
form quantum well states in this alternating potential, and are localized.
The $s$ states, much lighter by comparison, are simply scattered
but do not localize.

\section{Conclusions}

We studied an idealized Ni monatomic wire, and discovered that a domain
wall in that wire has a very important blocking effect, selectively
suppressing  the free motion of the $d$ electrons, but not of the 
$s$ electrons.
At a nanocontact separating two tips with reversed magnetizations,
the effective potential offset can act as a strong barrier yielding
total reflection for the heavy $d$ electrons alone. It is believed 
that this effect may be at work in some of the data quoted,
and it will be interesting to explore further consequences of this
mechanism experimentally.

Acknowledgments

This work was sponsored by MURST COFIN99, by INFM/F, PRA NANORUB,  
INFM/G, and "Iniziativa Transversale calcolo parallelo".
Calculations have been performed on the CRAY T3E at CINECA in Bologna using
the PWSCF package 
\cite{pw}.

\vfill
\eject

\begin{figure}
\caption{ 
Total energy of nonmagnetic and ferromagnetic Ni monatomic wires
as a function of interatomic spacing. Inset: magnetic moment per atom
of the ferromagnetic Ni wire. Bulk and atomic values are shown 
by dashed lines.}
\label{fig1}
\end{figure}

\begin{figure}
\caption{ 
Band structures of the nonmagnetic and ferromagnetic monatomic Ni wires 
at optimized interatomic spacing. The points where the bands cross
the Fermi level are marked by circles and the corresponding conducting
channels are numbered.}
\label{fig2}
\end{figure}

\begin{figure}
\caption{ 
Monatomic Ni wire with $M=0$; a) planar profile
of the magnetization along the wire. Note the formation of
sharp domain walls; b) band structure with the 
band indices and the conducting channel indication. Note 
that some $d$ bands have turned to flat levels.} 
\label{fig3}
\end{figure}

\begin{figure}
\caption{
Iso-electron density surfaces (4 levels from 0.0003 a.u.$^{-3}$ to 0.03 a.u.$^{-3}$)
for a) the conducting $s-d_{z^2}$ state and 
b) the localized $d_{xy},d_{x^2-y^2}$ states. All the states are calculated at 
$K=0.04~(2\pi/8a)$.  The lateral view (in the $yz$ plane) and the
cross section (in the $xy$ plane) at the $z$ coordinate indicated by the 
dashed line are shown. The  $d_{xy},d_{x^2-y^2}$
electrons are clearly reflected by the domain wall, and here form a 
quantum well state between two consecutive walls.
}
\label{fig4}
\end{figure}


\begin{thebibliography}{12}
\bibitem [8] {ref} The core radii (in a.u.) of our pseudopotentials are:
$r^c_s=(2.0,2.5)$, $r^c_p=(2.4,2.6)$, $r^c_d=(1.6,2.5)$.
Two values of the core radii indicate a channel which has been
pseudized within the ultrasoft scheme. In such case the first value
is the norm conserving core radius and the second is the ultrasoft 
one.
\end{thebibliography}
\end{document}